\newcommand{\CLASSINPUTtoptextmargin}{0.75in}%
\newcommand{\CLASSINPUTbottomtextmargin}{1in}%
\newcommand{\CLASSINPUTinnersidemargin}{0.63in}%
\newcommand{\CLASSINPUToutersidemargin}{0.63in}%
\documentclass[conference,10pt,letterpaper]{IEEEtran}%
\usepackage{amsmath}
\usepackage{graphicx}
\usepackage{multirow}
\usepackage[none]{hyphenat}
\usepackage{float}
\usepackage{subfig}
\usepackage{dblfloatfix}
\usepackage{booktabs}
\usepackage{eso-pic}
\usepackage{t1enc}
\usepackage{times}

\makeatletter

\def\@IMSauthorblockNAMEstyle{\normalfont\IMSauthorsize}
\def\@IMSauthorblockAFFILstyle{\normalfont\IMSaffilsize}
\def\@IMSauthorblockEMAILstyle{\normalfont\IMSaffilsize}
\def\IMSauthorblockNAME#1{%
\relax\@IMSauthorblockNAMEstyle%
#1%
}%
\def\IMSauthorblockAFFIL#1{%
\relax\@IMSauthorblockAFFILstyle%
\vskip\@IEEEauthorblockAtopspace
#1%
}%
\def\IMSauthorblockEMAIL#1{%
\relax\@IMSauthorblockEMAILstyle%
\vskip\@IEEEauthorblockAtopspace
#1%
}%
\newcommand{\IMSauthor}[1]{%
\ifIsBlindReviewVersion%
\author{\phantom{\parbox{\textwidth}{\center\relax#1}}}%
\else%
\author{\parbox{\textwidth}{\center\relax#1}}%
\fi%
}%
\newif\ifIsBlindReviewVersion

\def\IMSthispaperforfinalpublication{\IsBlindReviewVersionfalse}
\def\@maketitle{\newpage
\bgroup\par\addvspace{0.5\baselineskip}\centering%
\ifCLASSOPTIONtechnote
   {\bfseries\large\@IEEEcompsoconly{\sffamily}\@title\par}\vskip 1.3em{\lineskip .5em\@IEEEcompsoconly{\sffamily}\@author
   \@IEEEspecialpapernotice\par{\@IEEEcompsoconly{\vskip 1.5em\relax
   \@IEEEtitleabstractindextextbox{\@IEEEtitleabstractindextext}\par
   \hfill\@IEEEcompsocdiamondline\hfill\hbox{}\par}}}\relax
\else
   \vskip0.2em{\IMStitlesize\ifCLASSOPTIONtransmag\bfseries\LARGE\fi\@IEEEcompsoconly{\sffamily}\@IEEEcompsocconfonly{\normalfont\normalsize\vskip 2\@IEEEnormalsizeunitybaselineskip
   \bfseries\Large}\@title\par}\vskip1.0em\par
   \ifCLASSOPTIONconference%
      {\@IEEEspecialpapernotice\mbox{}\vskip\@IEEEauthorblockconfadjspace%
       \mbox{}\hfill\begin{@IEEEauthorhalign}\@author\end{@IEEEauthorhalign}\hfill\mbox{}\par}\relax
   \else
      \ifCLASSOPTIONpeerreviewca
         {\@IEEEcompsoconly{\sffamily}\@IEEEspecialpapernotice\mbox{}\vskip\@IEEEauthorblockconfadjspace%
          \mbox{}\hfill\begin{@IEEEauthorhalign}\@author\end{@IEEEauthorhalign}\hfill\mbox{}\par
          {\@IEEEcompsoconly{\vskip 1.5em\relax
           \@IEEEtitleabstractindextextbox{\@IEEEtitleabstractindextext}\par\hfill
           \@IEEEcompsocdiamondline\hfill\hbox{}\par}}}\relax
      \else
         \ifCLASSOPTIONtransmag
           {\@IEEEspecialpapernotice\mbox{}\vskip\@IEEEauthorblockconfadjspace%
            \mbox{}\hfill\begin{@IEEEauthorhalign}\@author\end{@IEEEauthorhalign}\hfill\mbox{}\par
           {\vspace{0.5\baselineskip}\relax\@IEEEtitleabstractindextextbox{\@IEEEtitleabstractindextext}\vspace{-1\baselineskip}\par}}\relax
         \else
           {\lineskip.5em\@IEEEcompsoconly{\sffamily}\sublargesize\@author\@IEEEspecialpapernotice\par
           {\@IEEEcompsoconly{\vskip 1.5em\relax
            \@IEEEtitleabstractindextextbox{\@IEEEtitleabstractindextext}\par\hfill
            \@IEEEcompsocdiamondline\hfill\hbox{}\par}}}\relax
         \fi
      \fi
   \fi
\fi\par\addvspace{0.0\baselineskip}\egroup}

\def\IMStitlesize{\@setfontsize{\IMStitlesize}{18}{21pt}}
\def\IMSauthorsize{\@setfontsize{\IMSauthorsize}{12}{13pt}}
\def\IMSaffilsize{\@setfontsize{\IMSaffilsize}{12}{13pt}}
\def\IMScaptionsize{\@setfontsize{\IMScaptionsize}{8}{9pt}}
\def\IMSbibsize{\@setfontsize{\IMSbibsize}{8}{9pt}}

\def\@IEEEauthorblockNstyle{\IMSauthorsize\@IEEEcompsocnotconfonly{\sffamily}\@IEEEcompsocconfonly{\large}}
\def\@IEEEauthorblockAstyle{\IMSaffilsize\@IEEEcompsocnotconfonly{\sffamily}\@IEEEcompsocconfonly{\itshape}\@IEEEcompsocconfonly{\large}}
\def\@IEEEauthordefaulttextstyle{\IMSauthorsize\@IEEEcompsocnotconfonly{\sffamily}\sublargesize}

\def\thebibliography#1{\section*{\refname}%
    \addcontentsline{toc}{section}{\refname}%
    \IMSbibsize\@IEEEcompsocconfonly{\small}\vskip 0.3\baselineskip plus 0.1\baselineskip minus 0.1\baselineskip
    \list{\@biblabel{\@arabic\c@enumiv}}%
    {\settowidth\labelwidth{\@biblabel{#1}}%
    \leftmargin\labelwidth
    \advance\leftmargin\labelsep\relax
    \itemsep \IEEEbibitemsep\relax
    \usecounter{enumiv}%
    \let\p@enumiv\@empty
    \renewcommand\theenumiv{\@arabic\c@enumiv}}%
    \let\@IEEElatexbibitem\bibitem%
    \def\bibitem{\@IEEEbibitemprefix\@IEEElatexbibitem}%
\def\newblock{\hskip .11em plus .33em minus .07em}%
\ifCLASSOPTIONtechnote\sloppy\clubpenalty4000\widowpenalty4000\interlinepenalty100%
\else\sloppy\clubpenalty4000\widowpenalty4000\interlinepenalty500\fi%
    \sfcode`\.=1000\relax}

\long\def\@makecaption#1#2{%
\ifx\@captype\@IEEEtablestring%
\par\@IEEEtabletopskipstrut
\else
\@IEEEfigurecaptionsepspace
\fi
\setbox\@tempboxa\hbox{\normalfont\IMScaptionsize {#1.}\nobreakspace\nobreakspace #2}%
\ifdim \wd\@tempboxa >\hsize%
\setbox\@tempboxa\hbox{\normalfont\IMScaptionsize {#1.}\nobreakspace\nobreakspace}%
\parbox[t]{\hsize}{\normalfont\IMScaptionsize\noindent\unhbox\@tempboxa#2}%
\else
\ifCLASSOPTIONconference \hbox to\hsize{\normalfont\IMScaptionsize\hfil\box\@tempboxa\hfil}%
\else \hbox to\hsize{\normalfont\IMScaptionsize\box\@tempboxa\hfil}%
\fi\fi
\ifx\@captype\@IEEEtablestring%
\@IEEEtablecaptionsepspace
\else
\fi}

\newlength\tablecaptiontotableskip
\newlength\figuretocaptionskip
\def\@IEEEfigurecaptionsepspace{\vskip\figuretocaptionskip\relax}%
\def\@IEEEtablecaptionsepspace{\vskip\tablecaptiontotableskip\relax}%

\def\abstract{\normalfont%
\@IEEEabskeysecsize\bfseries\textit{\abstractname}\,\bfseries\textit{---}\,%
\@IEEEgobbleleadPARNLSP}%

\def\IEEEkeywords{\normalfont%
\@IEEEabskeysecsize\bfseries\textit{\IEEEkeywordsname}\,\bfseries\textit{---}\,%
\@IEEEgobbleleadPARNLSP}%
\def\endIEEEkeywords{\relax\vspace{0.67ex}%
\par\if@twocolumn\else\endquotation\fi%
\normalsize\normalfont}%

\def\@IEEEauthorblockNtopspace{0ex}
\def\@IEEEauthorblockAtopspace{1mm}
\def\tablename{Table}
\def\thetable{\arabic{table}}
\def\IEEEkeywordsname{Keywords}
\def\subsubsection{\@startsection{subsubsection}{3}{\z@}{1.5ex plus 1.5ex minus 0.5ex}%
{0.7ex plus .5ex minus 0ex}{\normalfont\normalsize\itshape}}%
\def\@seccntformat#1{\csname the#1dis\endcsname\relax}
\def\thesectiondis{\thesection.\hskip 0.5em}
\def\thesubsectiondis{{\hbox to\parindent{\Alph{subsection}.}}}
\def\thesubsubsectiondis{{\hbox to \parindent{\arabic{subsubsection})}}}
\def\theparagraphdis{{\hbox to \parindent{\alph{paragraph})}}}
\IEEEilabelindentA \parindent
\IEEEilabelindent \IEEEilabelindentA
\IEEEelabelindent \parindent
\IEEEdlabelindent \parindent
\IEEElabelindent \parindent

\newlength\@IMSparindent
\newcommand\IMSdisplayacksection[1]{%
\ifIsBlindReviewVersion%
\noindent\phantom{\parbox[t]{\columnwidth}{\normalbaselines\setlength{\parindent}{\@IMSparindent}{#1}\strut}}
\else%
\noindent\parbox[t]{\columnwidth}{\normalbaselines\setlength{\parindent}{\@IMSparindent}{#1}\strut}%
\fi%
}%

\makeatother

\begin{document}
\raggedbottom
%
%
%
\title{Deep Learning-Driven Inverse Design of Doherty Power Amplifiers Using Pixelated Combiners and Dual-State Impedance Synthesis}


%
%
%
\IMSthispaperforfinalpublication
\IMSauthor{%
\IMSauthorblockNAME{
Han Zhou\textsuperscript{\protect\#\$}, Haojie Chang\textsuperscript{\protect\$}, David Widén\textsuperscript{\protect\$},
Christian Fager\textsuperscript{\protect\$}
}
\\%
\IMSauthorblockAFFIL{\textsuperscript{\protect\#}Tampere University, Tampere, Finland}
\\%

\IMSauthorblockAFFIL{\textsuperscript{\protect\$}Chalmers University of Technology, Gothenburg, Sweden
}
\\%
\IMSauthorblockEMAIL{
han.zhou@ieee.org\\
}
}
%

%
\AddToShipoutPictureFG*{%
  \AtPageUpperLeft{%
    \hspace{0.63in}%
    \raisebox{-0.32in}{%
      \parbox{\dimexpr\paperwidth-1.26in\relax}{%
        \centering\footnotesize\itshape
        This is the author-accepted version of a paper accepted for presentation at the
        2026 IEEE/MTT-S International Microwave Symposium (IMS 2026).
        \copyright~2026 IEEE. The published version is available at https://doi.org/10.1109/IMSRFTT43070.2026.11602564
      }%
    }%
  }%
}
\maketitle


%
%
%
\begin{abstract}
The output combiner of a Doherty power amplifier (PA) integrates load modulation, impedance matching, and phase compensation within a single network, making its design and synthesis highly challenging. In this paper, we propose a three-port Doherty combiner design methodology that combines deep convolutional neural networks (CNNs), pixelated layout representations, and genetic algorithms (GA) with dual-state impedance synthesis to address both peak and back-off power conditions. As a proof of concept, two GaN HEMT Doherty PA prototypes incorporating three-port pixelated combiners are designed and fabricated. Both prototypes achieve a measured saturated output power exceeding $44.2~\mathrm{dBm}$ with peak drain efficiency above $71.2\%$ within 2.6--2.8~GHz. Furthermore, a drain efficiency as high as $64\%$ is measured at the $6$-dB back-off level. After applying digital predistortion, each prototype achieves an adjacent channel leakage ratio (ACLR) better than $-51.3~\mathrm{dBc}$.


\end{abstract}

%
%
\section{Introduction}
The growing demand for higher data throughput in modern communication systems has driven the adoption of complex modulation schemes, resulting in signals with high peak-to-average power ratios (PAPR). To efficiently transmit these high-PAPR signals, power amplifiers (PA) must employ efficiency enhancement techniques such as load modulation. Among the various load-modulated PA architectures proposed both historically~\cite{DohertyPA, OutphasingPA} and in recent years~\cite{LMBA1, LMBA2, SLMBA1, SLMBA2, CLMA2}, Doherty PAs remain one of the most attractive solutions due to their simple implementation, decent linearity, and excellent efficiency under back-off conditions.

A critical aspect of Doherty PA design is the synthesis of its output combiner network. This network must simultaneously provide essential functionality such as load modulation, impedance transformation, and phase compensation, while also maintaining low loss and compact size to achieve high efficiency and enable miniaturization. The traditional Doherty design approach relies on parameterized models with pre-selected topologies composed of lumped or distributed components, such as transmission lines, inductors, and capacitors, followed by optimization through electromagnetic (EM) parameter sweeps. While this approach can be effective, it is often iterative, time-consuming, and prone to converging on local optima.

In recent years, deep learning–based inverse design approaches using pixelated EM structures have gained traction in radio frequency (RF) circuit design, demonstrating potential for expanding the design space and reducing design time~\cite{DL_PA_JSSC}. However, these methods have so far been largely limited to relatively simple PA architectures, such as wideband class~B PAs~\cite{DL_PA_JSSC}, harmonic-tuned class F/F\textsuperscript{-1} PAs~\cite{AI_HZ, AI_F-1_HZ}, and template-based multi-layer differential PAs~\cite{DL_6GPA} etc. Applying deep learning techniques with pixelated layouts to the design of Doherty combiners has remained unexplored, owing to the inherent complexity of synthesizing three-port load-modulated networks. Nevertheless, this capability is highly desirable, as the combiner fundamentally governs Doherty PA behavior and ultimately sets the limits on achievable back-off efficiency.

In this paper, we introduce a dual-state impedance synthesis framework for efficient Doherty combiner design. A deep convolutional neural network (CNN)-based surrogate model is developed to capture the nonlinear relationships between pixelated layout structures and frequency-dependent S-parameters. Combined with a genetic algorithm (GA) for dual-state impedance optimization, this approach bridges the gap between deep learning-driven pixelated design and Doherty combiner synthesis, enabling rapid design space exploration and efficient Doherty network realization. The proposed methodology is validated through two GaN HEMT Doherty PA prototypes employing pixelated combiners, both achieving excellent measured performance.

\section{Deep Learning-Driven Combiner Synthesis}
In this study, circuit layouts are represented as binary matrices, where "1" indicates the presence of a metal pixel and "0" non-metal. The design area is discretized into a $15 \times 15$ grid, each cell sized $1.8\,\mathrm{mm} \times 1.8\,\mathrm{mm}$ (a pixel), with element size expanded by 20\% to allow diagonal paths. Exhaustive exploration of all pixel combinations would require $2^{225}$ possibilities, making brute-force search infeasible. We, therefore, train a deep CNN based on pixelated layouts and their corresponding S-parameters to act as a fast EM surrogate model, delivering sub-millisecond predictions. Integrated with a GA, it enables rapid design-space exploration and generation of pixelated layouts that meet Doherty combiner specifications.

\begin{figure*} [t!]
    \centering     
    \includegraphics[width=0.91\textwidth]{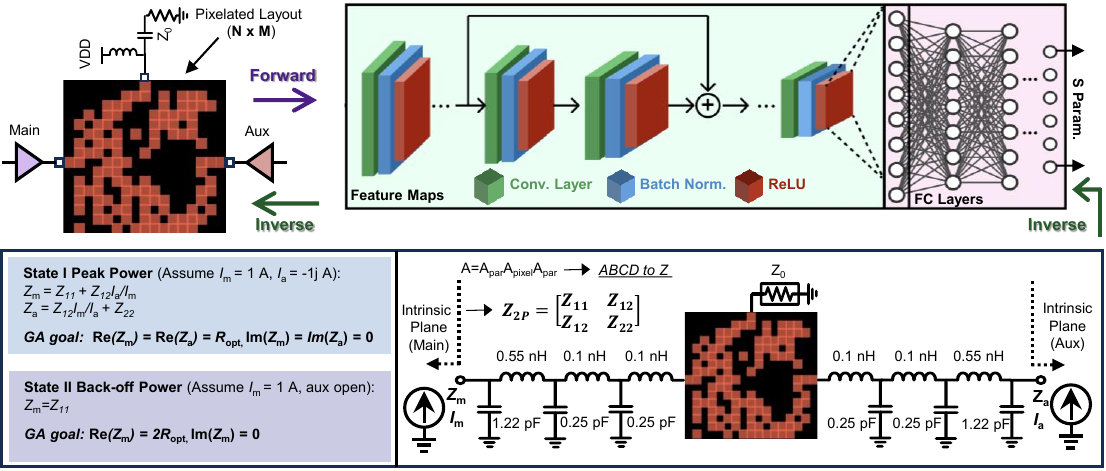}
    \caption{Inverse design workflow for pixelated Doherty combiners using deep convolutional neural networks (CNN) and the proposed dual-state impedance synthesis approach integrated with genetic algorithms (GA).}
    \label{fig.1}
\end{figure*}

\subsection{Data Generation, Model Architecture, and Training Details}
To train the neural network, we first generate a dataset of approximately 75K circuits, later augmented to 600K samples. Each training circuit is created by assigning a random metal coverage to the layout, drawn from a normal distribution centered at 50\% with a 15\% standard deviation. The initial data is then obtained through EM simulations in ADS Momentum, using frequencies within $2.4-3.0$~GHz with 0.05~GHz spacing. The layout includes four ports positioned at the center of each side, and during augmentation, one port is terminated with an open circuit while another is terminated with a 50~$\Omega$ load.

The employed deep CNN architecture is illustrated in Fig.~\ref{fig.1}. It begins with an input layer for the circuit matrices, followed by 12 convolutional blocks. Each block consists of 32 convolutional filters, and batch normalization layers. To address the vanishing gradient problem and improve stability, we incorporate a deep residual network structure~\cite{He_2016_CVPR}. After the convolutional layers, six fully connected (FC) layers are employed, each with 2048 neurons and dropout applied at a rate of 25\% to prevent overfitting. Both convolutional and dense layers use Leaky ReLU (LeReLU) activations, enabling the network to learn complex nonlinear relationships. The output layer predicts the S-parameters, including real and imaginary parts, across 13 discrete frequencies within the 2.4--3.0~GHz range. The network is trained for 300 epochs using the mean absolute error (MAE) loss function and the Adam optimizer with a learning rate of 0.001.

\begin{figure*} [t!]
    \centering     
    \includegraphics[width=0.95\textwidth]{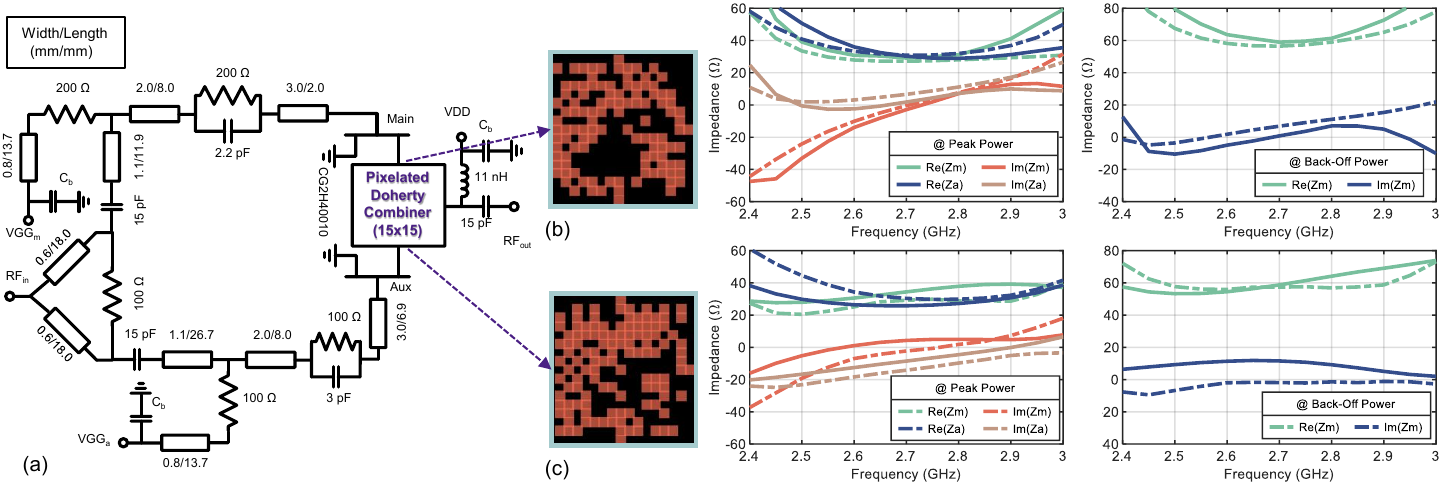}
    \caption{(a) Circuit schematic of the proposed Doherty PA prototype with two synthesized combiners: Combiner 1 (b) and Combiner 2 (c), along with CNN-based emulator evaluation comparing predicted impedances (dashed) at the main and auxiliary current-source planes for peak and back-off power against full-wave EM simulated results (solid) from ADS Momentum.}
    \label{fig.2}
\end{figure*}

\begin{figure*} [t!]
    \centering     
    \includegraphics[width=0.95\textwidth]{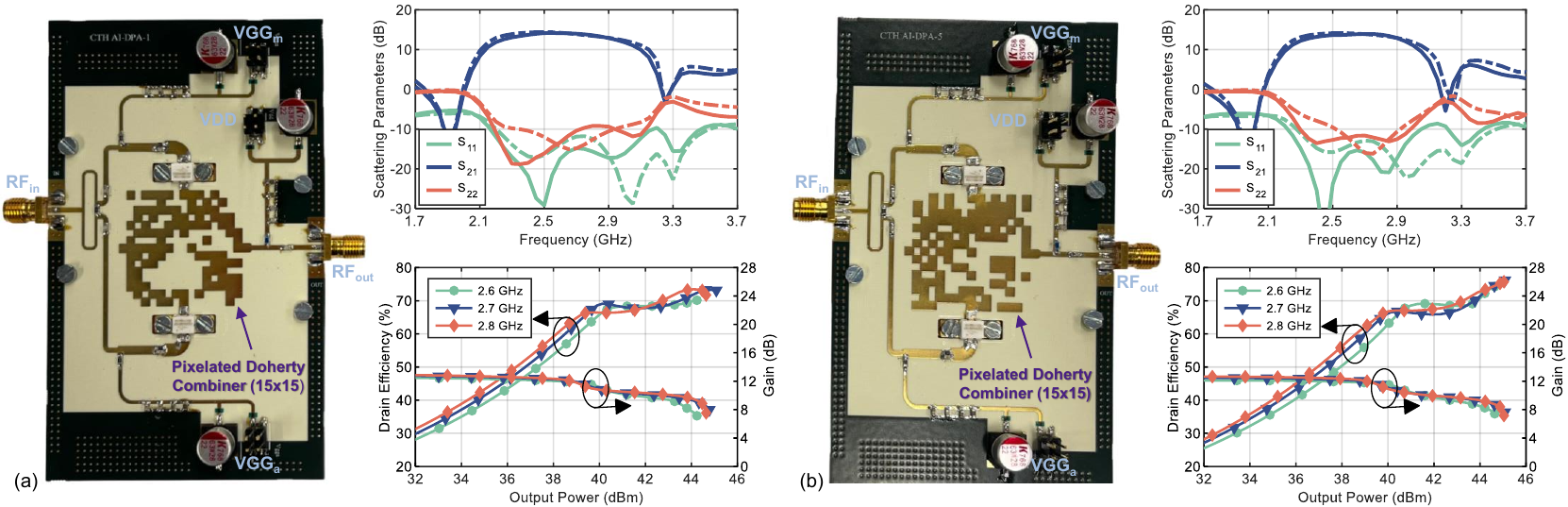}
    \caption{Two fabricated Doherty PAs with Combiner 1 (a) and Combiner 2 (b), showing simulated (dashed) and measured (solid) small-signal results, along with large-signal continuous-wave (CW) measurements over 2.6--2.8~GHz.}
    \label{fig.3}
\end{figure*}

\subsection{Dual-State Doherty Combiner Synthesis}
We use a GA to synthesize pixelated EM layouts based on target specifications. An initial population of binary matrices is generated, and in each generation, a CNN surrogate predicts S-parameters while fitness scores quantify deviation from the target. The top 10 individuals are retained as elites, while up to 30\% of the population is replaced with random layouts. Parents are selected via tournament, and mutation flips pixels to introduce variability. This process evolves the population toward layouts that meet the desired performance. 

A key challenge in Doherty combiner synthesis is defining targets that accurately represent Doherty operation. To address this, we propose a dual-state impedance synthesis method. The combiner is modeled as a three-port network: ports 1 and 2 connect to the main and auxiliary amplifiers, while port 3 is terminated with a $50~\Omega$ load. Transistor parasitics and packaging elements are de-embedded and incorporated following the approach in~\cite{AI_HZ}. The overall network is represented by cascaded ABCD matrices, which are then converted into an impedance matrix $Z_{2\mathrm{P}}$, representing the pixelated combiner with parasitics. The impedance at the main and auxiliary current-source planes is expressed as

\begin{equation}
\displaystyle Z_{\mathrm{m}} = Z_{11} + \displaystyle Z_{12}\frac{I_{\mathrm{a}}}{I_{\mathrm{m}}}
\label{eq:1}
\end{equation}
\begin{equation}
\displaystyle Z_{\mathrm{a}} = Z_{22} + \displaystyle Z_{12}\frac{I_{\mathrm{m}}}{I_{\mathrm{a}}}
\end{equation}

For different back-off levels, the ratio $I_{\mathrm{a}}/I_{\mathrm{m}}$ varies. In this work, we consider a symmetrical Doherty configuration with two operating states: 

\begin{itemize}
  \item \textbf{State I (Peak Power):} 
  At peak power, $|I_{\mathrm{a}}| = |I_{\mathrm{m}}|$ and $\angle(I_{\mathrm{a}}, I_{\mathrm{m}}) = 90^\circ$, so
  \[
  I_{\mathrm{a}} = -j I_{\mathrm{m}} 
  \]

  \item \textbf{State II (Back-Off Power):} 
  At back-off, the auxiliary PA is off ($I_{\mathrm{a}} = 0$), giving
  \begin{equation}
  Z_{\mathrm{m}} = Z_{11}
  \label{eq:Zbo}
  \end{equation}
\end{itemize}

To synthesize the pixelated Doherty combiner, we first compute the $Z_{2\text{P}}$ parameters. With three unknowns and three equations (\ref{eq:1})--(\ref{eq:Zbo}), the system can be solved. At peak power, the target impedance for both main and auxiliary amplifiers is~$R_{\mathrm{opt}}=(V_{\mathrm{BR}}-V_{\mathrm{knee}})/I_{\mathrm{max}}$
and at back-off power, the main amplifier’s impedance becomes $2R_{\text{opt}}$. Our GA goal is thus to ensure that, within $2.6-2.8$~GHz band: \begin{itemize}
    \item $\Re(Z_{\mathrm{m,peak}})$ \& $\Re(Z_{\mathrm{a,peak}})$ remain within $\pm 10\%$ of $R_{\text{opt}}$,
    \item $\Re(Z_{\mathrm{m,bo}})$ remains within $\pm 10\%$ of $2R_{\text{opt}}$,
\end{itemize}
while minimizing the imaginary components.

\section{Doherty PA Prototype Circuit Design}
To validate the proposed inverse design methodology, we developed two Doherty PA prototypes using pixelated combiners synthesized by the deep learning approach. The complete schematics are shown in Fig.~\ref{fig.2}(a). Both prototypes share the same circuit topology except for the combiners, demonstrating the diversity and robustness of the synthesis method. The designs are implemented on a 20-mil Rogers 4350B substrate using 10-W packaged GaN HEMT transistors (Macom CG2H40010F) for both main and auxiliary devices. As shown in Fig.~\ref{fig.1}, the employed transistor's output parasitics and packaging elements are well characterized, and its estimated optimal load is $R_{\text{opt}} = 30~\Omega$~\cite{Ropt}. 

Fig.~\ref{fig.2}(b) and (c) present CNN-based emulator evaluations comparing predicted Doherty impedances at the main and auxiliary current-source planes for peak and back-off power against full-wave EM simulated results from ADS Momentum. While minor discrepancies exist in the predicted impedances, its profile closely matches EM results across 2.6--2.8~GHz.


\section{Measurement Results}
The photograph of the two fabricated Doherty PAs is shown in Fig.~\ref{fig.3}. During measurement, for both prototypes, we conduct three types of tests: small-signal, large-signal with continuous-wave (CW), and modulated signal with digital predistortion (DPD). The drain bias is set to $V_{\mathrm{DD}} = 28~\mathrm{V}$. The main amplifier is biased with a quiescent current of $40~\mathrm{mA}$, while the auxiliary gate bias is fixed at $-7~\mathrm{V}$.

\subsection{Small-Siganl and Large-Siganl CW Measurement}
The proper response of both Doherty prototypes is confirmed by the good agreement between S-parameter simulations and measurement results, as shown in Fig.~\ref{fig.3}. Across the 2.2--3.0~GHz band, the measured small-signal gain ranges from 11.8--14.3~dB for prototype~1 and 10.1--13.9~dB for prototype~2. The measured input return loss remains below -11.9~dB and -11.4~dB for prototype~1 and~2, respectively. The CW results, including drain efficiency and gain versus power over 2.6--2.8~GHz, are depicted in Fig.~\ref{fig.3}. Clearly, distinct Doherty profiles with significant efficiency enhancement are observed for both circuits. Within the 2.6--2.8~GHz band, the measured peak drain efficiency ranges from $71.2$--$73.1\%$ for prototype~1 and $75.2$--$76.5\%$ for prototype~2. The measured back-off efficiency is $57.3$--$64.6\%$ and $56.3$--$63.8\%$ for prototype~1 and prototype~2, respectively. Furthermore, the measured output power exhibits $44.2$--$44.8$ and $44.3$--$44.7$~dBm for prototype 1 and 2, respectively.

\subsection{Modulated-Signal Measurement }

\begin{figure} [t!]
    \centering     
    \includegraphics[width=\columnwidth]{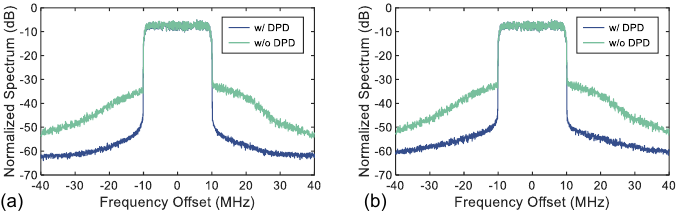}
    \caption{Normalized spectrum of Doherty prototypes 1 (a) and 2 (b) measured under 20~MHz 7~dB PAPR OFDM signals at 2.7~GHz with and without DPD.}
    \label{fig.4}
\end{figure}

The performance of both prototypes are also examined using a $20$-MHz orthogonal frequency division multiplexing (OFDM) signal with a $7$-dB PAPR signal. As shown in Fig.~\ref{fig.4}, the adjacent channel leakage ratio (ACLR) improves from $-29.8$ to $-52.7$~dBc for prototype~1 and from $-31.2$ to $-51.3$~dBc for prototype~2 after applying DPD~\cite{ILC}.

As summarized in Table~\ref{tab.1}, the two fabricated Doherty PAs demonstrate excellent CW performance compared to state-of-the-art designs, along with strong linearizability under modulated signal testing.

\begin{table}[t]
    \centering
    \caption{Summary of State-of-the-art Load-modulated PAs.}
    \begin{tabular}{ c c c c c c c }
    \toprule  
    \multirow{2}{*}{ Ref.} & \multirow{2}{*}{Arch.} & Freq & $\eta$\textsubscript{SAT} & $\eta$\textsubscript{BO-6dB} & P\textsubscript{SAT} & ACLR\\
    & & (GHz) & ($\%$) & ($\%$) & (dBm) & (dBc) \\
    \midrule
    \multirow{1}{*}{ \cite{T1}'25} & 3-DPA & \multirow{1}{*}{0.7--0.8} & 56--62 & 38--49 & 43 & \multirow{1}{*}{N.A.} \\
    \midrule
    \multirow{1}{*}{ \cite{Table2}'23} & 2-DPA& \multirow{1}{*}{3.3--3.9} & 48--53 & 34--45 &  45.6 & \multirow{1}{*}{N.A.} \\
    \midrule
    \multirow{1}{*}{ \cite{DPA_kenle}'20} & 2-DPA& \multirow{1}{*}{3.4-3.6} & 63-70 & 44-55 &  41.8 & \multirow{1}{*}{N.A.} \\
    \midrule
    \multirow{1}{*}{ \cite{Table4}'22} & LMBA & \multirow{1}{*}{$2.4$} & $54$ & $47$ &  $44.1$& \multirow{1}{*}{-48.0} \\
    \midrule
    \multirow{1}{*}{ \cite{RFinCLMA}'24} & CLMA& \multirow{1}{*}{3.3--3.5} & 52--57& 51--53 & 42.2& \multirow{1}{*}{-51.6} \\
    \midrule
    \multirow{2}{*}{\textbf{This Work}} & \textbf{DPA 1} & 2.6--2.8 & 71--73 & 57--64 & 44.2 & -52.7 \\
     & \textbf{DPA 2} & 2.6--2.8 & 75--76 & 56--64 & 44.3 & -51.3 \\
 
    \bottomrule    
    \end{tabular}\\
    \label{tab.1}
\end{table}
\section{Conclusion}
We propose a dual-state impedance synthesis approach that integrates deep CNNs, pixelated layout representations, and GAs for the design of three-port Doherty combiners. Two GaN Doherty PA prototypes are implemented, achieving peak drain efficiency above $71.2\%$ and saturated output power exceeding $44.2~\mathrm{dBm}$ within the 2.6--2.8~GHz band. Furthermore, a drain efficiency of up to $64\%$ is measured at the $6$-dB back-off level. The proposed deep learning-driven design methodology offers a scalable and efficient solution for advancing multi-port load-modulation RF PA architectures.

\section*{Acknowledgment}
This research was supported in part by Swedish Innovation Agency (VINNOVA), Sivers Semiconductors, and Chalmers University of Technology under Grant 2022-00863, and in part by VINNOVA Grant 2024-02531, MULTIRACS, through the Eureka CELTIC Framework.

\newcommand{\IMSacktext}{%
Authors wish to acknowledge...
}



\bibliographystyle{IEEEtran}

\bibliography{IEEEabrv,IEEEexample}

\end{document}